\documentclass{article}

\usepackage{tcolorbox}
\usepackage{graphicx}   
                        
\usepackage[T1]{fontenc}
\usepackage{listings}   
\usepackage{setspace}   
\usepackage{hyperref}   
\usepackage{todonotes}

\makeindex              

\begin{document}
\doublespacing
\date{}
\title{Economical Accommodations for Neurodivergent Students in Software Engineering Education: Experiences from an Intervention in Four Undergraduate Courses}
\author{Grischa Liebel and Steinunn Gróa Sigurðardóttir}
%
%
\maketitle

\section{Introduction}
Neurodiversity is an umbrella term that describes variation in brain function among individuals \cite{armstrong2015myth}, including conditions such as Autism spectrum disorder (ASD), Attention deficit hyperactivity disorder (ADHD), or dyslexia.
%
Neurodiversity is common in the general population, with an estimated 5.0\% to 7.1\% \cite{willcutt2012prevalence,polanczyk2007worldwide} and 7\% \cite{peterson2012developmental} of the world population being diagnosed with ADHD and dyslexia respectively.
Neurodivergent (ND) individuals often experience challenges in specific tasks, such as difficulties in communication or a reduced attention span in comparison to neurotypical (NT) individuals \cite{ucu_workplace}.
However, they also exhibit specific strengths, such as high creativity \cite{white2011creative} or attention to detail \cite{baron2009talent}.
Therefore, improving the inclusion of ND individuals is desirable for economic, ethical, and for talent reasons.

In higher education, struggles of ND students are well-documented \cite{griffin2009student,gelbar2014systematic,pino2014inclusion}.
Common issues in this area are a lack of awareness among other students and staff, forms of assessment that are particularly challenging for some students, and a lack of offered accommodations.
These factors commonly lead to stress, anxiety, and ultimately a risk of dropping out of the studies.

Accommodations for ND students can require substantial effort.
However, smaller changes in course material can already have major impact.
In this chapter, we summarise the lessons learned from an intervention in four courses in undergraduate computer science programmes at Reykjavik University, Iceland, over a period of two terms.
Following accessibility guidelines produced by interest groups for different ND conditions, we created course material in the form of slides and assignments specifically tailored to ND audiences.
We focused on small, economical changes that could be replicated by educators with a minimal investment of time.
We evaluated the success of our intervention through two surveys, showing an overall positive response among ND students and neurotypical (NT) students.
Example materials we produced are available at \url{https://doi.org/10.5281/zenodo.7199162}.

\begin{tcolorbox}
\textbf{In summary, in this chapter you will learn:}
\begin{enumerate}
    \item Which obstacles ND students commonly face in higher education.
    \item What accommodations are recommended for ND students.
    \item What economical interventions for Computer Science students can look like.
    \item What feedback we received on making such an intervention.
\end{enumerate}
\end{tcolorbox}

\section{The State of Knowledge in Research}
Work on neurodiversity is broad and covers different areas.
In addition to work within the realms of Medicine and Psychology, there are many studies investigating neurodiversity in higher education, neurodiversity and employment, and piloted interventions that aim to improve inclusion of ND students in education or ND individuals in the workplace.
We briefly summarise these three areas as follows.

\begin{tcolorbox}
\textbf{Neurodiversity in Higher Education and Industry}
\begin{enumerate}
    \item ND students often under-perform in specific task, due to the task's nature.
    \item Awareness and accommodations for ND students are commonly lacking.
    \item ND students often feel anxious, depressed, or afraid of being stigmatised.
    \item ND individuals in industry often choose to not disclose their condition.
    \item Accommodations for ND individuals in industry vary considerably.
    \item There is only initial work on neurodiversity in Software Engineering.
\end{enumerate}
\end{tcolorbox}

ND conditions are often diagnosed in early childhood.
Therefore, much of existing work targets primary and secondary school education.
As inclusion of ND students improves in these areas, the amount of ND students enrolling in higher education increases.
That is, educators in higher education need to increase their awareness of neurodiversity and possible accommodations for ND students \cite{jurecic2007neurodiversity}.
In particular, educators should understand that ND students make mistakes or under-perform in certain tasks due to the tasks' nature, not due to a lack of intelligence.

Experiences of ND students in higher education are published in several studies.
A summary of research themes surrounding neurodiversity in higher education is presented by Clouder et al.~\cite{clouder2020neurodiversity}, based on 48 screened publications.
The findings show that many higher education institutes provide insufficient support for neurodiversity, primarily due to ``low levels of staff awareness, ambivalence and inflexible teaching and assessment approaches''.
ND students often feel frustrated due to the lack of available accommodations, anxious or frightened due to challenging situations or stigmatisation, and often embarrassed to ask questions.
Finally, inclusive and trusting environments, as well as accommodations such as extended exam times or flexible assignment schedules benefit ND students.

Experiences and support of college and university students with autism are summarised in a systematic literature review (SLR) by Gelbar et al.~\cite{gelbar2014systematic}.
The authors emphasise the need for support in higher education, and that anxiety and depression among autistic students are common.

Regarding students with dyslexia in higher education, Pino and Mortari~\cite{pino2014inclusion} summarise existing research, finding that research is fragmented and contains many gaps.
The article reports coping strategies such as getting support from family and friends, or adapting their writing styles for written assignments.
Furthermore, the authors highlight a common lack of awareness and acceptance by academic staff, and mention that accommodations are usually appreciated by students.

In the IT industry, inclusion of ND individuals is typically at an early stage and companies need to become more welcoming and inclusive \cite{wille18}.
Similar to higher education, ND individuals often do not disclose their condition due to fear of stigmatisation \cite{morris15,lindsay19}.
However, depending on the workplace the number of ND individuals disclosing their condition and the offered accommodations differ considerably.

Driven by the Covid-19 pandemic, several studies investigate how remote work affects ND individuals.
Das et al.~\cite{das2021towards} find that ND individuals create accessible workplaces at home, negotiate communication and meeting practices, and balance tensions between productivity at work and fatigue.
The authors suggest to commonly record audio and video, enable automated meeting transcripts, and make meeting notes available routinely.
In a similar direction, the needs of adults with ASD in video calling are studied by Zolyomi et al.~\cite{zolyomi2019managing}.
The authors find that adults with ASD develop several coping strategies, such as adopting neurotypical behaviour, and that they experience more stress than NT individuals without adaptations.
The authors provide suggestions for adaptations, e.g., translating and communicating social and emotional information to adults with ASD.

Specific to Software Engineering, there is only initial work on neurodiversity and potential accommodations for ND individuals.
Morris et al.~\cite{morris15} investigate challenges faced by ND individuals.
Specific themes related to SE were the tendency to get bored with mundane tasks, or expressing inappropriate emotions, e.g., when criticised during code reviews.
Furthermore, ND individuals perceive several tasks more challenging than NT individuals, e.g., working in shared offices or noisy settings, or deciding when to seek help for tasks.
Begel et al.~\cite{begel2021remote} report the results of a successful 13-day game programming camp for incoming college students with ASD.
In addition to teaching game development, students were instructed specifically in communication skills.
In a line of research investigating how dyslexic individuals read and comprehend software code, McChesney and Bond~\cite{mcchesney18,mcchesney2018b,mcchesney20,mcchesney21,mcchesney21b} find that, contrary to intuition, dyslexic individuals do not exhibit over-proportional deficiencies compared to NT individuals.
Potential explanations are that reading code is significantly different from regular text, e.g., due to indentation or spacing.

\section*{Possible Accommodations for Neurodivergent Students}
Several guidelines on how to accommodate ND students have been published by interest groups for different neurodivergent conditions.
These guidelines typically target one specific condition at a time.
Deciding on possible accommodations for a class or a university therefore requires synthesising these guidelines into one set of concrete accommodations that somehow align well with several existing guidelines.
This is a difficult step, given that existing guidelines might contradict each other and might propose too many accommodations for a feasible intervention in practice.
In the following we will summarise three guidelines, one for each of the most common ND conditions ASD, ADHD, and dyslexia.

\begin{tcolorbox}
\textbf{Accommodation Areas for Different ND Conditions}
\begin{enumerate}
    \item ASD: Sensory environment, communication, escape ways, awareness.
    \item ADHD: Patience and understanding, clear structure, communication.
    \item Dyslexia: Style and presentation of visual material.
\end{enumerate}
\end{tcolorbox}

For ASD, we consider a checklist for Autism-friendly environments published by the NHS\footnote{\url{https://positiveaboutautism.co.uk/uploads/9/7/4/5/97454370/checklist_for_autism-friendly_environments_-september_2016.pdf}}.
The checklist describes accommodations to four areas: the sensory environment, e.g., reducing strong colours or overly patterned surfaces, or reducing smell and noise; to the communication environment, e.g., the use of clear and unambiguous signs; to provide escape ways for autistic individuals in case of high stress levels; and to provide general awareness of ASD among the employees. 

For ADHD, we considered a report by the ADHD foundation on how to teach and manage students with ADHD\footnote{\url{https://www.adhdfoundation.org.uk/wp-content/uploads/2022/03/Teaching-and-Managing-Students-with-ADHD.pdf}}.
The report provides a general introduction to ADHD and to common behaviours of individuals with ADHD.
Then, suggestions are made on how classrooms can be made more accessible to students with ADHD.
First, teachers are encouraged to show patience and understanding for behaviour that might seem odd, such as excessive movement or inappropriate comments.
Second, providing a structure and flexibility is important, e.g., through regular routines, or by providing overviews before starting a class or a checklist for an assignment.
Finally, the report lists a few hints on communicating with students with ADHD, e.g., trying to tell the student what they should do instead of what not to do, or addressing them by name.

Finally, for dyslexia, we consider the style guide by the British Dyslexia Association\footnote{\url{https://www.bdadyslexia.org.uk/advice/employers/creating-a-dyslexia-friendly-workplace/dyslexia-friendly-style-guide}}.
This guide focuses on accommodations in the style of visual material, e.g., assignment texts or lecture slides.
Some of the changes they recommend are using fonts that are easier to read for dyslexic individuals, e.g., Arial and Comic Sans\footnote{Yes, you read that correctly.}; increasing font size; increasing spacing between letters, words and lines; using left-aligned instead of justified text; avoiding multi-column texts; and using cream-coloured backgrounds instead of plain white.
Additionally, adapting the writing style is recommended, e.g., by avoiding passive voice, using concise sentences, or avoiding jargon.
Many of these recommendations overlap with traditional style guidelines for presentations, e.g., using large font sizes and avoiding red and green as colours, and for academic writing, e.g., avoiding overly long sentences and jargon.

\section{An Example Intervention in Four Courses} 
Based on the guidelines and the existing academic work outlined in the previous section, we designed an intervention in initially three courses which we were teaching in the fall term 2021 at Reykjavik University, Iceland.
In the spring term 2022, we used the intervention in a slightly altered way in one more course.
All four courses are part of the undergraduate computer science and software engineering programmes in the first and second year at Reykjavik University.
The courses are an introductory Software Engineering course, an introductory web development course, and two courses on discrete mathematics.
During the interventions, we were operating in a hybrid mode of on-site teaching and remote teaching through video conferencing tools.
Prior to our intervention, the only accommodations ND students could receive were extended exam times, usually limited to the final exams only.
Approximately 10\% of the students at Reykjavik University have a diagnosed ND condition and applied for this extension, though the percentage was higher in some of the four courses.

The accommodations were made as a part of the regular teaching.
That is, there was no specific time, budget, or workforce assigned to making the chosen accommodations.
Therefore, we focused on small, economical changes that we could implement under regular time pressure.
In summary, we made the following changes.
\begin{tcolorbox}
\textbf{Implemented Accommodations}
\begin{enumerate}
    \item Emphasis on style changes in assignments and lecture slides.
    \item Improved clarity and structure.
    \item Minor changes in sensory environment, e.g., consistent lighting.
    \item Discussing neurodiversity and the accommodations to raise awareness.
\end{enumerate}
\end{tcolorbox}
Specifically, we decided to primarily change the style of our assignments and slides.
We changed font type to OpenDyslexic\footnote{\url{https://opendyslexic.org}}, the background colour to a cream colour, increased inter-letter, intra-letter, and line spacing, avoided italics, multiple columns, and specific colours, such as red and green, or strong colours in general.
We did not change the default fonts in case of mathematical formulas and program code.
To leave the choice of style up to the students, we offered assignments both in a traditional style and in our updated, ND-friendly style.

In terms of clarity and communication, we added bullet-point summaries to all assignments.
Additionally, we proof-read the assignments and tried to reduce ambiguity, jargon and unclear acronyms, and improve overall clarity.
Examples of the slides and assignments created in this way can be found at \url{https://doi.org/10.5281/zenodo.7199162}.

In addition to adapting the style and content of our slides and assignments, we made minor changes to the sensory environment, e.g., avoiding clothes with strong patterns or colours, and in remote teaching using high-quality microphones and consistent artificial lighting.
Finally, we discussed the changes and the reasons behind them in the classroom, with the intention to raise awareness for neurodiversity and make ND students feel seen and welcome.

Overall, the accommodations were efficient.
For assignments, we used LaTeX and thus only had to re-compile the assignment with slightly changed parameters.
Re-reading assignments and slides and improving clarity was done as a part of updating existing assignments, which we argue should be part of preparing assignments in any case.
Changing the sensory environment primarily required us to be conscious about potential issues, such as avoiding clothes with strong patterns or colours.
Finally, the largest effort was required for updating lecture slides in Microsoft PowerPoint.
While we made style changes on the slide master level, i.e., globally for all slides, increased font size and spacing required changes to individual slides to fit all content.
To update existing slides took approximately 30 minutes per slide deck (per lecture).
This effort is likely much lower if you create lecture slides from scratch, and not, as in our case, update existing ones.

Through two surveys, one after each term, we evaluated the impact our intervention had on all students in our courses.
We received answers from 169 students in all courses, corresponding to an overall response rate of 23.57\%.
Of those who answered, we only considered those that completed the survey, resulting in 155 valid answers.
Out of the 155 students, 63 (40.64\%) self-identified as neurodivergent, 89 as neurotypical, and 3 did not disclose.
This corresponds to 108 students (48 ND, 59 NT, 1 not answered) in the Fall term 2021 (3 courses), and 47 students (15 ND, 30 NT, 2 not answered) in the Spring term 2022 (1 course).

\section{Intervention Experiences}
\begin{figure*}[ht]
    \centering
     \includegraphics[width=1\linewidth]{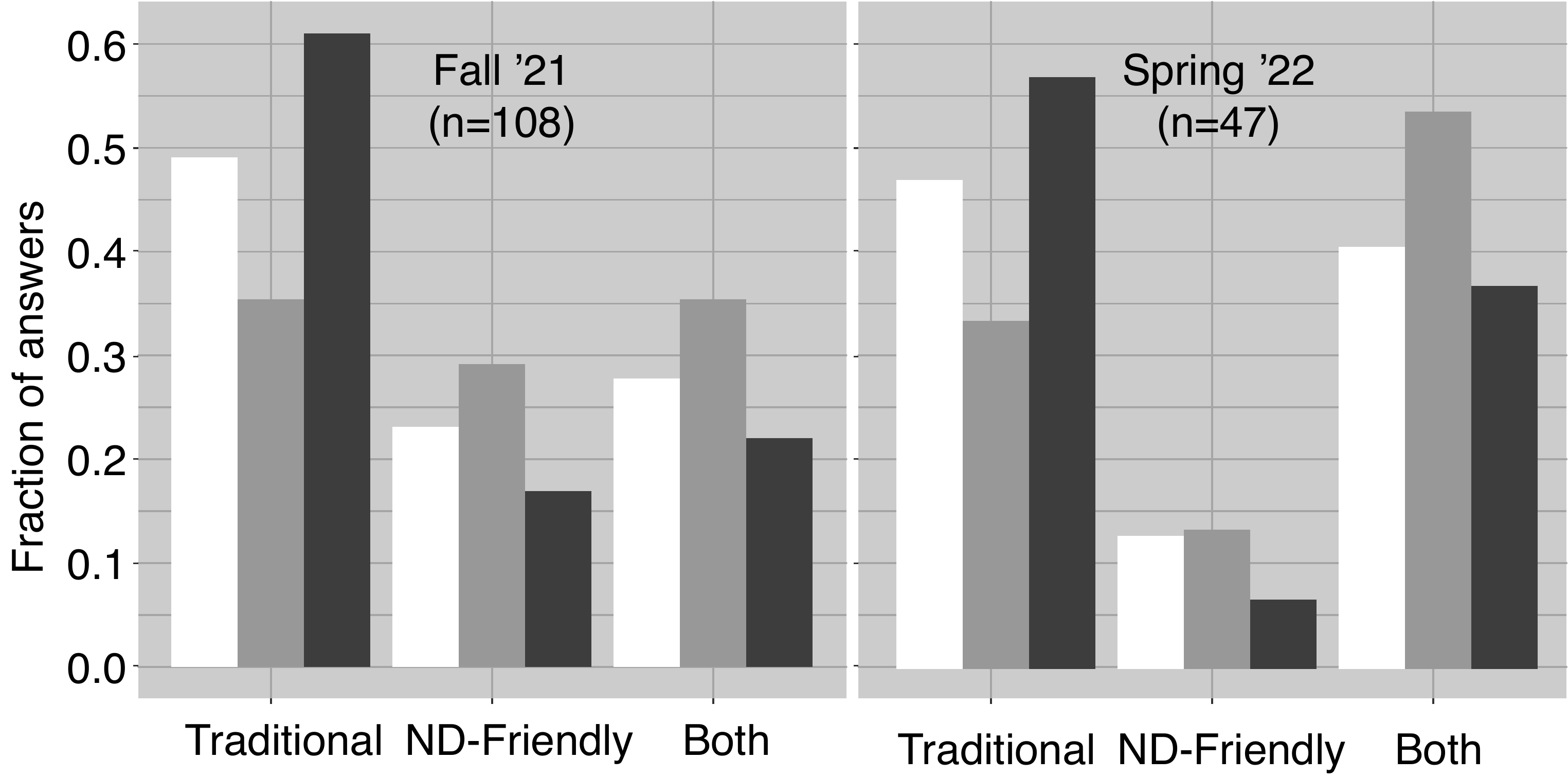}
	\caption{Used materials in the fall term 2021 (n=108) and spring term 2022. All students are depicted in the white bars, while ND students are shown in grey and NT students in black.}
	\label{fig:sur_material_comp}
\end{figure*}

While our students had the choice which material they used, ND-friendly material was used by over 50\% of the respondents, among them many NT students.
The detailed break down is depicted in Figure~\ref{fig:sur_material_comp} for the three courses in the fall term 2021, and for the remaining course in the spring term 2022.
Interestingly, several NT students commented in free text answers that they also found the ND-friendly material more readable compared to the traditional style and therefore preferred it.

Many students, both NT and ND, further expressed gratitude that we raised awareness for neurodiversity.
Finally, five ND students stated that they felt welcome in our courses.
Given that academic literature regularly points out that ND students experience unwelcoming environments in higher education, we consider this one of the most important outcomes of our intervention.

We further asked participants to rate whether they (a) appreciate that material is provided in two styles, (b) find the ND-friendly material harder to read, and (c) find that the ND-friendly material helps them understand the content better.
The results are depicted in Table~\ref{tab:perceptions}.
Note that the statements in the first column are shorter, simplified versions of the actual statements, which can be found in the material we provide.
\begin{table}[ht]  
  \caption{Perceptions by All Participants Regarding the ND-friendly Course Material in Fall and Spring Terms.}
  \smallskip
  \label{tab:perceptions} 
  \centering
  \begin{tabular}{p{8.8cm}p{2.6cm}p{2.6cm}}
    \hline\noalign{\smallskip}
    \textbf{Statement} & \textbf{Fall '21} & \textbf{Spring '22}\\
     & \textbf{(n=108)} & \textbf{(n=47)}\\
    \hline\noalign{\smallskip}
    I appreciate that material is provided in two styles. & 1\% disagree,\newline 5\% neutral,\newline \textbf{94\% agree} & 4\% disagree,\newline 4\% neutral,\newline \textbf{91\% agree}\\
    \hline\noalign{\smallskip}
    The ND-friendly material is harder to read. & \textbf{65\% disagree},\newline 13\% neutral,\newline 22\% agree & \textbf{52\% disagree},\newline 8\% neutral,\newline \textbf{40\% agree}\\
    \hline\noalign{\smallskip}
    The ND-friendly material helps me understand the content better.  & 27\% disagree,\newline 30\% neutral,\newline 42\% agree & 34\% disagree,\newline 31\% neutral,\newline 34\% agree\\
    \hline\noalign{\smallskip}
  \end{tabular}
\end{table}
The answers show that accommodations were received well by most of the participants.
Furthermore, a large proportion of all participants found that the ND-friendly material helped them understand the material better.
Considering only ND participants, the agreements are naturally higher.
That is, 48\% of the ND students in the first survey agree that the ND-friendly material helped their understanding (compared to 34\% of the NT students), and 54\% of the ND students agree in the second survey (compared to only 12\% of NT students).
This shows that the accommodations reached their goal of primarily supporting ND students.
Nevertheless, the divided agreements and disagreements also clearly show that our simple accommodations are no silver bullet.

%
In addition to closed questions, we received a lot of noteworthy feedback as free-text answers.
Four students found the changed font particularly hard to read, especially in assignments, and therefore preferred the traditional material, even though it might not be ideal either.
For example, one NT student commented in free text that they felt ``very distracted'' by the new font. 
Given this feedback and the fact that we used a non-standard font, we are currently considering to use a standard font that is considered suitable for dyslexic audiences in the future, e.g., Arial.
Three ND students commented that they did not want accommodations, or that they were satisfied with extended exam time already provided by the university.
Finally, a few comments showed the individual nature of neurodiversity.
One dyslexic student commented on the difficulty of following a course offered in English only, while another student expressed difficulties as the course they attended was given in Icelandic, but the course book was English.
That is, they struggled primarily with the multi-lingual nature of the course.
Similarly, one ND student stated that working from home had a positive impact on their productivity, while another student stated during one of our courses that they found it especially hard to focus at home due to their ND condition.
Overall, these comments show that simple changes are unlikely to cater to all ND audiences, and that flexibility in accommodations is needed.
However, it is worth pointing out that we did not receive any negative feedback.

In addition to direct feedback on our accommodations, several students provided suggestions on additional accommodations.
Particularly common were suggestions to provide video or audio recordings in which we explain assignments, or simply read them out loud.
This clearly highlights difficulties many ND students face with text comprehension.
However, it also shows the potential benefit of text-to-speech tools that have been improving substantially in the last years.
Another common point for improvement is the organisation and presentation of a single course.
Several ND students pointed out that it can be hard to find all the relevant information on a course.
Additionally, syllabi in our programmes are currently not standardised, which leads to large differences in how courses present information such as learning outcomes.
This poses an additional obstacle for many ND students.
We believe these two directions are alternatives to our style and presentation based approach that could provide particularly valuable accommodations, i.e., different modalities for course material and improved organisation.

\section{Summary}
In this chapter, we summarised potential accommodations for neurodivergent (ND) students in higher education, focusing on the three most common conditions Autism Spectrum Disorder (ASD), Attention Deficit Hyperactivity Disorder (ADHD), and dyslexia.
We outlined which accommodations we made to four courses in Computer Science and Software Engineering undergraduate programs, focusing on changes in style to assignments and slides that are easy to implement with limited time.
Our results show that minor accommodations for ND students can have an important impact, from some ND students feeling more seen and welcome in courses, to actual perceived improvements in understanding the course material.
Researchers and educators can use our provided material to conduct studies and develop their own interventions.

Our accommodations were of general nature, even though we applied them in a computer science and Software Engineering context.
We consider this an important starting point before applying accommodations that target specific characteristics of or tasks in Software Engineering.
Future work in this area should target how to convey course organisation in a better way, and how to use other media than text for expressing tasks, assignments, and project descriptions, such as video or audio recordings.
Furthermore, we believe that more work is needed that studies what tasks in Software Engineering are particularly difficult or easy for ND individuals, and why.
To this end, we have started investigating which strengths ND individuals exhibit with respect to Software Engineering tasks, and how they can use these in a better way.

\begin{tcolorbox}
\textbf{Take-Away Points}
\begin{enumerate}
    \item Neurodiversity is not rare, and is increasing in higher education.
    \item Many accommodations are difficult and effort intensive.
    \item Small, economical changes can make a difference.
    \item Awareness is a first, important step.
    \item Feedback clearly shows the individual nature of neurodiversity.
\end{enumerate}
\end{tcolorbox}

%


\bibliographystyle{plain}
\bibliography{sample}
\end{document}